\documentclass[10pt,a4paper]{article}
\usepackage{jcappub}

\usepackage{graphicx,color,amsmath,amsxtra,bm}
\usepackage{amssymb}
\usepackage[english]{babel}
\usepackage{amsfonts}
\usepackage{dcolumn}
\usepackage{subfigure}
\usepackage{ifthen}

\newcommand{\apj}{{\em Astrophys. J. }}
\newcommand{\nat}{{\em Nature }}
\newcommand{\prd}{{\em Phys. Rev. }{\bf D }}
\newcommand{\prl}{{\em Phys. Rev. Lett. }}

\newcommand{\mnras}{{\em Mon. Not. Roy. Astron. Soc. }}
\newcommand{\apjl}{{\em Astrophys. J. Let. }}
\newcommand{\apjs}{{\em Astrophys. J. Suppl. }}
\newcommand{\jcap}{{\em JCAP }}
\newcommand{\physrep}{{\em Phys. Rept. }}

\newcommand{\aap}{{\em Astron. Astrophys. }}
\newcommand{\aj}{{\em Astron. J. }}
\newcommand{\pasj}{{\em Publ. Astron. Soc. Japan }}

\begin{document}

{\baselineskip0pt
\rightline{\baselineskip16pt\rm\vbox to20pt{
            \hbox{YITP-20-74}
\vss}}%
}

\title{New measures to test modified gravity cosmologies}
	
\author[a,b,1]{Jiro Matsumoto,\note{Corresponding author.}} \emailAdd{ushirugapen@gmail.com}	
\author[c,d]{Teppei Okumura,} \emailAdd{tokumura@asiaa.sinica.edu.tw}
\author[b,d,e]{and Misao Sasaki} \emailAdd{misao.sasaki@ipmu.jp}

\affiliation[a]{Wind Energy Institute of Tokyo, Inc., Kawate Bld. Suite 501, 1-5-8 Nishi-Shinbashi, Minato-Ku, Tokyo 105-0003, Japan}
\affiliation[b]{LeCosPA, National Taiwan University, Taipei 10617, Taiwan }
\affiliation[c]{Academia Sinica Institute of Astronomy and Astrophysics (ASIAA), No. 1, Section 4, Roosevelt Road, Taipei 10617, Taiwan}
\affiliation[d]{Kavli Institute for the Physics and Mathematics of the Universe (WPI), UTIAS, The University of Tokyo, Kashiwa, Chiba 277-8583, Japan}
\affiliation[e]{Yukawa Institute for Theoretical Physics, Kyoto University, Kyoto 606-8502, Japan}
  
\abstract{
The observed accelerated expansion of the Universe may be explained by dark energy or the breakdown of general relativity (GR) on cosmological scales.
When the latter case, a modified gravity scenario, is considered, it is often assumed that the background evolution is the same as the $\Lambda$CDM model but the density perturbation evolves differently. In this paper, we investigate more general classes of modified gravity, where both the background and perturbation evolutions are deviated from those in the $\Lambda$CDM model. We introduce two phase diagrams, $\alpha{\rm-}f\sigma _8$ and $H{\rm-}f\sigma _8$ diagrams; $H$ is the expansion rate, $f\sigma_8$ is a combination of the growth rate of the Universe and the normalization of the density fluctuation which is directly constrained by redshift-space distortions, and $\alpha$ is a parameter which characterizes the deviation of gravity from GR and can be probed by gravitational lensing. We consider several specific examples of Horndeski's theory, which is a general scalar-tensor theory, and demonstrate how deviations from the $\Lambda$CDM model appears in the $\alpha{\rm-}f\sigma _8$ and $H{\rm-}f\sigma _8$ diagrams.
The predicted deviations will be useful for future large-scale structure observations to exclude 
some of the modified gravity models.
}
\arxivnumber{2005.09227}
\maketitle

\section{Introduction \label{sec1}}
The accelerated expansion of the current Universe has been clarified by the observations of type Ia supernovae in the late 1990s \cite{Riess:1998,Perlmutter:1999} 
and supported by observations of cosmic microwave background radiation (CMB) \cite{Komatsu:2011,Planck-Collaboration:2014,Planck-Collaboration:2016}
and baryon acoustic oscillations (BAO) \cite{Eisenstein:2005,Okumura:2008,Percival:2010,Blake:2011b,Beutler:2011,Cuesta:2016,Delubac:2015}. 
Broadly speaking, there are two approaches to explain the accelerating Universe; (i) introducing an additional energy component, called dark energy, to the total energy budget and (ii) modifying the action of gravity from the one predicted by Einstein's theory of relativity. 
In dark energy models, one adopts the Einstein-Hilbert action and introduces a fluid matter with the equation-of-state parameter $w< -1/3$, where $w =-1$ corresponds to a cosmological constant. 
The simplest dark energy model is $\Lambda$CDM, and more generally, the $w$CDM or quintessence model has been considered \cite{Peebles:1988,Ratra:1988,Chiba:1997,Zlatev:1999}. 
On the other hand, the Einstein-Hilbert action itself is modified in the models of modified gravity theories. 
The models include $F(R)$ gravity \cite{Buchdahl:1970,Nojiri:2006,Sotiriou:2010,De_Felice:2010,Nojiri:2011,Nojiri:2017}, 
massive gravity \cite{de_Rham:2011,Hassan:2012,Hassan:2012a}, 
and Horndeski's theory \cite{Horndeski:1974}. 

The differences of the models appear both in the background equations and in the linear perturbation equations. 
While $\Lambda$CDM is consistent with most of the observations, 
a tension between the values of the Hubble constant, $H_0$, constrained from early universe and late universe 
started to be recognized after  the Planck mission reported the first results \cite{Planck-Collaboration:2014}. 
The descrepancy in the $H_0$ values between local observarions and CMB observations was first reported using Cepheid variables \cite{Marra:2013,Riess:2016,Di_Valentino:2016}, and it has also been seen 
in the observations of strong lensing time delay and so on \cite{Wong:2019,Chen:2019,Birrer:2019,Jee:2019,Shajib:2020,Verde:2019}. 
Moreover, there is another tension in a parameter, $f\sigma_8$, where $f$ is the growth rate of the universe and $\sigma_8$ is the normalization of the density fluctuation amplitude.
 This parameter can be directly constrained through peculiar velocities of galaxies in galaxy redshift surveys, known as redshift-space distortions (RSD) \cite{Kaiser:1987,Hamilton:1998}, and is used to distinguish modified gravity models \cite{Linder:2005,Jain:2008,Guzzo:2008,Song:2009,Okumura:2011}. However, the $f\sigma_8$ values observed so far at $z<1$ are systematically lower than the prediction of the best-fitting $\Lambda$CDM model from the Planck result \cite{Blake:2011,Samushia:2012,de_la_Torre:2013,Reid:2014,Song:2015,Okumura:2016}, as pointed out by, e.g., Refs. \cite{Kazantzidis:2018,Nesseris:2017}. 
Therefore, the importance of reconsidering the dynamics of the Universe in modified gravity models is
increasing. 

Modified gravity theories, in general, have too many degrees of freedom to be completely analyzed. 
Consequently, the observables of perturbation quantities (e.g. growth rate of the matter density perturbation)
have been investigated only for simple models or for phenomenologically parametrized models. 
Moreover, the background evolution in modified gravity models is often assumed to be the same as 
that in the $\Lambda$CDM model.
In this paper we take account of both the variations of background dynamics and those 
of perturbation quantities. For this purpose, we introduce new phase diagrams;
 the $\alpha{\rm-}f\sigma _8$ and $H{\rm-}f\sigma _8$ diagrams.
Similar diagrams can be found in Refs.~\cite{Perenon:2017,Linder:2017,Basilakos:2017,Linder:2018,Linder:2019}.
Here $\alpha$ describes the effect on gravitational lensing and is defined in Eq. (\ref{deflection})
 below in terms of the deflection potential $\Psi_{\rm defl}$, and $\alpha=0$ 
 in the case of general relativity
  (e.g., \cite{Koyama:2006a,Thomas:2009,Daniel:2010,Tereno:2011,Simpson:2013a,Matsumoto:2019}). 
Thus, these diagrams will enable us to investigate how each of the two key observations 
of large-scale structures, RSD and gravitational lensing, can constrain a given modified gravity model. 

In this paper, we consider Horndeski's theory, which is a general scalar-tensor theory.
We do not adopt phenomenological parameterizations which have been commonly used
in the literature. We focus several specific examples of Horndeski's gravity and 
demonstrate how deviations from the $\Lambda$CDM model appear in the $\alpha{\rm-}f\sigma _8$
 and $H{\rm-}f\sigma _8$ diagrams.

The contents of the paper are as follows. 
We briefly overview how models of dark energy and modified gravity behave 
on the $\alpha{\rm-}f\sigma _8$ and $H{\rm-}f\sigma _8$ diagrams in Sec.~\ref{sec2}.
For this purpose we present the former and latter diagrams for the simple representative cases 
of dark energy and modified gravity,  the $w$CDM and $F(R)$ gravity models, respectively. 
Then we describe Horndeski's theory in Sec.~\ref{sec3},
 and considering its typical examples, we present their $\alpha{\rm-}f\sigma _8$ and $H{\rm-}f\sigma _8$ 
diagrams in Sec.~\ref{sec4}. 
Our concluding remarks are given in Sec.~\ref{sec5}. 
A more detailed description of Horndeski's theory is presented in Appendix \ref{Htheory}. 
A short note on the possibility of having negative $\alpha$ is given in Appendix \ref{app}.

Throughout this paper, we adopt Natural units, $\hbar =c = k_{B}=1$, 
and the gravitational constant $8 \pi G$ is denoted by
${\kappa}^2 \equiv 8\pi/{M_{Pl}}^2$ 
with the Planck mass of $M_{Pl} = G^{-1/2} = 1.2 \times 10^{19}$GeV.

\section{Dark energy and modified gravity\label{sec2}}
In models in which dark energy is given by a matter field, commonly its energy momentum tensor 
is decoupled from the other matter components. In this case, the main effect of dark energy is to
modify the background evolution of the Universe, and hence its effect to the growth of linear 
perturbations is indirect.
On the other hand, in modified gravity theories of dark energy, its effective energy momentum tensor
is likely to be coupled to the matter components.
As a result, not only the background evolution of the Universe but also the linear perturbation equations 
may significantly deviate from those in the $\Lambda$CDM model.

In the following, we assume the metric,
\begin{align}
	ds^2 = -(1+2 \Psi (t,x) )dt^2 + a^2(t)(1+2\Phi (t,x)) \delta _{ij} dx^i dx^j\,,
\end{align}
and adopt notations in Fourier space as
\begin{align}
\Psi (t,k) + (1+\alpha (t,k)) \Phi (t,k)= 0 , \label{alens}\\
\delta \rho (t,k) / \rho (t) = \delta (t,k) = \hat \delta(k) D(t,k), \label{deltaD}
\end{align}
where 
$k$ is the wave number, $D(t,k)$ is  the growing mode of the matter density perturbation, 
and $\hat \delta (k)$ describes 
the scale dependence of $\delta (t,k) $ at initial time $t=t_i$ under the normalization of $D$ as $D(t_i,k)=1$. 

In Eq.~(\ref{alens}), we introduced a quantity $\alpha$, which vanishes in Einstein gravity (i.e., $\Psi+\Phi=0$). 
The relation of $\alpha$ with the well-known gravitational slip parameter, $\eta \equiv -\Phi / \Psi$ \cite{Jain:2008},
is $\eta = 1/(1+\alpha)$. 
Thus $\alpha$ characterizes how much a given gravity model deviates from General Relativity (GR). 
Because the strength of gravitational lensing is determined by the deflection potential \cite{Schimd:2005},
\begin{equation}
\Psi _{\rm defl} \equiv \Psi - \Phi =\frac{2+\alpha}{1+\alpha}\Psi
=\left(2-\frac{\alpha}{1+\alpha}\right)\Psi\,,
\label{deflection}
\end{equation}
we see that the gravitational lensing effect is reduced if $\alpha >0$ or $\alpha<-2$
and enhanced if $-2<\alpha <0$.
We note that there is a subtlety in this interpretation. It will be shown in section \ref{sec4} that,
for a class of modified gravity models considered in this paper, 
the lensing effect is unaffected if we compare it for the same mass distribution. 

To see this, let us first introduce $G_{\rm eff}$ which relates the Newton potential to the
matter density perturbation in modified gravity:
\begin{equation}
\Delta\Psi =4\pi G_{\rm eff}\delta\rho\,a^2\,.
\label{Geffdef}
\end{equation}
Let us also introduce $G_{\rm light}$ that would relate the deflection potential $\Psi_{\rm defl}$
to $\delta\rho$,
\begin{equation}
\Delta\Psi_{\rm defl} =4\pi G_{\rm light}\delta\rho\, a^2\,.
\label{Glightdef}
\end{equation}
We have $G_{\rm eff}=G$ and $G_{\rm light}=2G$ in GR. 
On the other hand, for modified gravity,
the $\alpha$-dependence of $G_{\rm eff}$ is found in section \ref{sec4} as 
\begin{equation}
G_{\rm eff}=\frac{2(1+\alpha)}{2+\alpha}G\,.
\label{Geffalp}
\end{equation}
From Eqs.~(\ref{deflection}) and (\ref{Geffdef}), we find
\begin{equation}
\Delta\Psi_{\rm defl} =\frac{2+\alpha}{1+\alpha}\Delta\Psi =\frac{2+\alpha}{1+\alpha}\cdot 4\pi G_{\rm eff}\delta\rho\,a^2\,.
\end{equation}
Hence we obtain
\begin{equation}
G_{\rm light}=\frac{2+\alpha}{1+\alpha}G_{\rm eff}\,.
\end{equation}
Inserting Eq.~(\ref{Geffalp}) into the above $G_{\rm eff}$ gives $G_{\rm light}=2G$,
that is, the resulting $\Psi _{\rm defl}$ will be the same as that in GR, independent of $\alpha$. 

However, as the mass distribution is measured by its gravitational effect that includes the
modification in the effective gravitational constant, what can be compared is the
lensing effect for the same $\Psi$. This means our original interpretation is correct from
an observational point of view. More precisely speaking, it is the gravitationally inferred mass
distribution that is overestimated if  $\alpha >0$ or $\alpha<-2$
and otherwise if $-2<\alpha <0$, while the gravitational lensing effect remains the same.

To quantify the growth of linear perturbations, one usually considers the combination, $f\sigma_8$,
inferred from galaxy surveys, where $f=d \ln D/ d \ln a$ and $\sigma_8$ is the normalization
 parameter of the density perturbation spectrum. 
In the following, to test modified gravity models, we evaluate the ratio of $f\sigma_8$ in a given model 
to that in the $\Lambda$CDM model, $f \sigma _8  / f \sigma _{8, \Lambda {\rm CDM}} $. 
Note that the value of $f$ should approach unity in any viable model of gravity in the high-redshift limit. 
We also assume that the value of $\sigma_8$ coincides with that of the $\Lambda$CDM model in the high redshift limit.  
Thus, this ratio becomes unity at high redshifts.
Since $f \sigma _8$ depends not only on the modifications of the linear perturbation equations 
but also on the background evolution of the Universe,  below
we will study correlations between $f \sigma _8$ and $\alpha$ and between
$f\sigma_8$ and the Hubble expansion rate $H$ in various models of gravity.
The evolution equation of the quasi-static mode of the matter density contrast $\delta = \delta \rho _m/ \rho _m$
is expressed as \cite{de_Felice:2011,Matsumoto:2019}
\begin{equation}
\ddot \delta + 2H \dot \delta - 4 \pi G_\mathrm{eff} \rho _m \delta = 0. 
\label{deltaeq}
\end{equation}
The parameter $f \sigma _8$ will be evaluated by using Eq.~(\ref{deltaeq}) with appropriate initial conditions. 

\begin{figure}
\begin{minipage}[t]{0.49\columnwidth}
\begin{center}
\includegraphics[width=0.97\columnwidth]{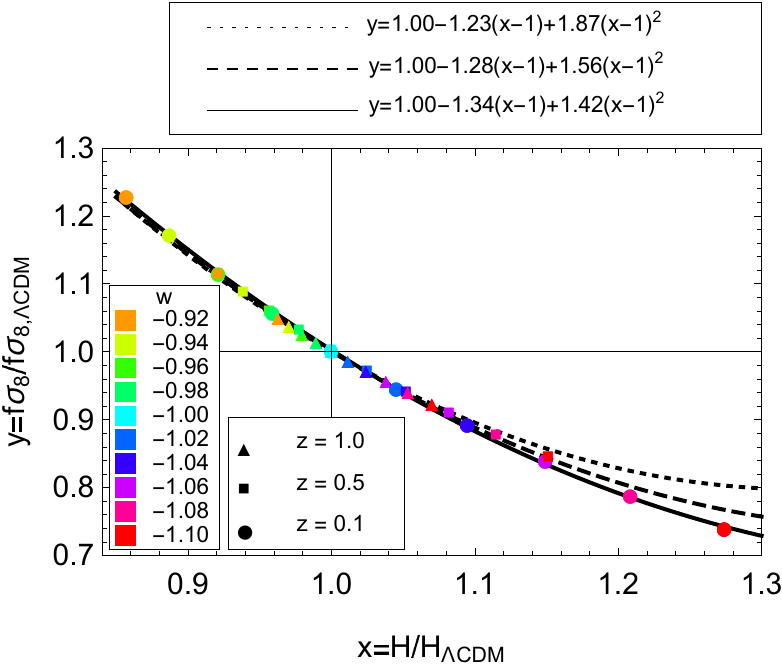}
\end{center}
\end{minipage}
\begin{minipage}[t]{0.49\columnwidth}
\begin{center}
\includegraphics[width=0.97\columnwidth]{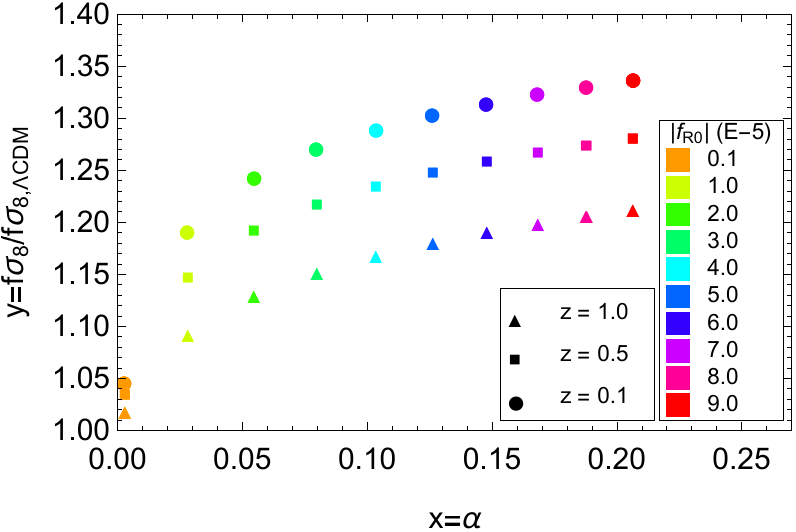}
\end{center}
\end{minipage}
\caption{(Left) $H{\rm-}f \sigma$ diagram for the $w$CDM model for 
 various values of the equation of state parameter $w$.
	Dark matter density is set to be the same as that in the $\Lambda$CDM model.
	Triangles, Squares, and Circles correspond to redshift $1.0$, $0.5$, and $0.1$, respectively. 
	In order to clarify the redshift dependences, the best-fitting quadratic curves are shown for the redshifts 1.0 (dotted), 0.5 (dashed) and 0.1 (solid).
	}
\label{wcdm}
\caption{(Right) $\alpha{\rm-}f \sigma_8$ diagram for the $F(R)$ gravity model with $F(R)=R -2 \Lambda +|f_{R0}| R_0^2/R$
	for various values of $|f_{R0}|$.}
\label{FR}
\end{figure}

Before presenting detailed studies on modified gravity models, let us first consider the
simplest dark energy model, i.e. the $w$CDM model. 
Note that $\alpha=0$
because there is no deviation from GR in this model.
The $H{\rm-}f \sigma _8$ diagram is depicted in Fig. \ref{wcdm}. 
The initial conditions are set at $z=10$ where they coincides with those of the $\Lambda$CDM model. 
The figure shows that the deviation in $f \sigma _8$ is almost proportional to that in the Hubble rate at each redshift,
with negative proportionality coefficients.
Thus $f \sigma _8$ in $w$CDM models becomes smaller than that in the $\Lambda$CDM model
for models with a larger Hubble rate $H$ which corresponds to $w<-1$. 
We note that the linear perturbation equations in the $w$CDM model are unchanged from those of 
the $\Lambda$CDM model.
Therefore, the smaller value of $f \sigma _8$ due to a larger $H$ may be regarded as a purely background effect. 
We should note that if we apply a different boundary condition, e.g., $H_{wCDM} = H_{\Lambda CDM}$ and 
$\Omega _{m} =0.31$ at $z=0$, the quantitative results will be different though the qualitative tendency will 
remain the same.


As we mentioned in the above, the behavior of the matter density perturbation is 
not independent from the background evolution of the Universe in general.
 In some theories, e.g. $F(R)$ gravity theories, even if the difference between the model and GR 
at the background level is negligibly small, there can be $O(1)$ difference at perturbation level
due to the scale dependence of the matter density perturbation that enhances the deviation from GR on small scales.
As a characteristic example for such a case, let us consider
$F(R)$ gravity with 
\begin{eqnarray}
F(R)=R -2 \Lambda +|F_{R0}|\frac{R_0^2}{R}\,,
\label{FRgrav}
\end{eqnarray}
where $R_0$ and $\Lambda$ are the current values of the Ricci scalar and the cosmological constant, respectively. 
For  the parameter $F_{R0}$ in the range $10^{-6} < | F_{R0}| < 10^{-4} \ll \Lambda /R_0$,
this model is known to reproduce the background evolution of the $\Lambda$CDM model almost exactly \cite{Hu:2007}.
In Fig.~\ref{FR}, we plot the predicted values of $\alpha$ and $f \sigma$ for several redshifts.
In general, the function $\delta (t,k) $ is not separable, in the other words, the $k$-dependence
of $D(t,k) $ in (\ref{deltaD}) cannot be ignored in $F(R)$ gravity theories. Here, for definiteness,
we plot $f\sigma_8 $, i.e. $f\sigma$ at $k=(8 h^{-1}$Mpc$)^{-1}$.
It shows that the density perturbation in this model always evolves faster than that in the case of the $\Lambda$CDM model,
and $\alpha$ is always positive.

\section{More general modified gravity models \label{sec3}}

In the previous section, we studied two examples; $w$CDM model and $F(R)$ gravity theory. 
In the $w$CDM case, the linear perturbation equations are the same as those in the $\Lambda$CDM model,
while the background evolution of the Universe is different.
On the other hand, in the $F(R)$ gravity case, the background evolution of the Universe is
the same as the $\Lambda$CDM model, while the linear perturbation equations are different.
In more general modified gravity theories, both of them will be different from the $\Lambda$CDM model. 
Here we consider Horndeski's theory. It is a general scalar-tensor theory which includes $F(R)$ gravity 
as a special case \cite{de_Felice:2011,OHanlon:1972,Chiba:2003}. 

The action in Horndeski's theory is given by \cite{Horndeski:1974,Deffayet:2011,Kobayashi:2011}
\begin{equation}
S_H=\sum ^5 _{i=2} \int d ^4 x \sqrt{-g} {\cal L}_i,
\end{equation}
where
\begin{align}
{\cal L}_2 &= K(\phi , X), \\
{\cal L}_3 &= -G_3(\phi , X) \Box \phi , \\
{\cal L}_4 &= G_4(\phi , X)R + G_{4 X} \left [ ( \Box \phi )^2 - (\nabla _\mu \nabla _\nu \phi)^2 \right ], \\
{\cal L}_5 &= G_5 (\phi , X)G_{\mu \nu} \nabla ^{\mu}\nabla ^{\nu} \phi - 
\frac{G_{5X}}{6} \left [ (\Box \phi)^3 - 3 (\Box \phi) (\nabla _ \mu \nabla _ \nu \phi)^2 +2 (\nabla _ \mu \nabla _ \nu \phi)^3 \right ]. 
\end{align}
Here, $K$, $G_3$, $G_4$, and $G_5$ are generic functions of 
$\phi$ and $X=- \partial _\mu \phi \partial ^\mu \phi /2$, and the subscript $X$ means a derivative with respect to $X$. 
The total action is the sum of $S_H$ and the matter action $S_\mathrm{matter}$ which contains baryons and cold dark matter. 
More details are referred to Appendix \ref{Htheory}

The recent observations of gravitational wave event GW170817 \cite{Abbott:2017}
and its electromagnetic counterparts \cite{Abbott:2017a,Abbott:2017b,Coulter:2017} 
showed that the propagation speed of gravitational waves should satisfy 
\begin{equation}
\vert c_T^2 -1 \vert \lesssim 10^{-15},
\end{equation}
in the relatively recent Universe. This bound implies the sound speed of the tensor mode,
given by
\begin{equation}
c_T ^2 = \frac{G_4 -XG_{5 \phi}-XG_{5X} \ddot \phi}{G_4 -2XG_{4X}-X(G_{5X}\dot \phi H - G_{5 \phi})}, 
\label{ct}
\end{equation}
should be almost unity, where and below the subscript $\phi$ means a derivative with respect to $\phi$. 
If the terms $XG_{5 \phi}$, $XG_{4X}$, $\cdots$  are relevant for the evolution of the Universe, then 
one expects a substantial deviation of $c_T^2$ from unity. 
Therefore, it is reasonable to assume that the terms proportional to $G_{4X}$, $G_{5 \phi}$, and $G_{5 X}$ are 
not relevant for the current accelerated expansion of the Universe. Hence we set $G_{4X}=G_5=0$ in the following.

\begin{figure}[tb]
\begin{center}
\includegraphics[width=0.5\columnwidth]{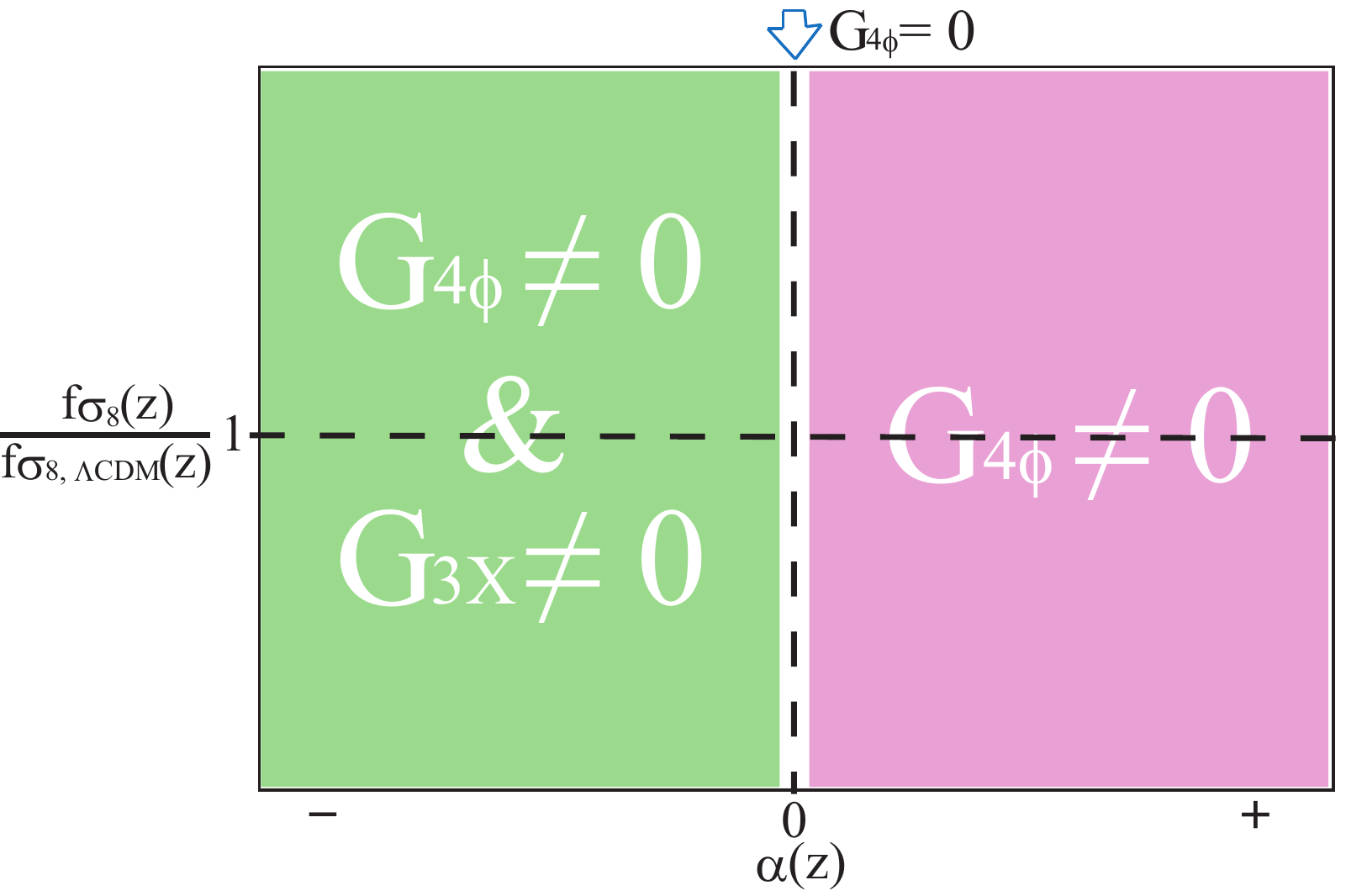}
\caption{Classification of Horndeski's models in terms of $\alpha$ and $f$ under the
	assumptions $K_X >0$ and $G_4 >0$. 
 Nonzero $\alpha$ is only realized when $G_{4\phi}\neq0$, and 
 $\alpha <0$ is realized only if  $G_{4 \phi} \neq 0$ and $G_{3X} \neq 0$. }
\label{region}
\end{center}
\end{figure}

Even after this simplification, the theory still remains quite complicated.
For example, the effective Newton constant that can be defined on sufficiently small scales
$k^2\gg a^2H^2$ takes the form~\cite{Matsumoto:2019},
\begin{eqnarray}
 G_{\rm eff}&=&\dfrac{1}{16 \pi G_4}\left(\dfrac{{\cal A}}{{\cal B}} +O(a^2H^2/k^2)\right)\,;
 \label{geff} \\
&{\cal A}&\equiv G_4 C_{\rm kin}+4 G_{4 \phi}^2 ,
 \label{geff1}\\
 &{\cal B}& \equiv G_4 C_{\rm kin}
 - \frac{1}{4} \dot \phi ^4 G_{3X}^2  
- \dot \phi ^2 G_{3X} G_{4 \phi} +3 G_{4 \phi} ^2 ,
 \label{geff2}
\end{eqnarray}
where we have introduced an effective coefficient of the kinetic term $C_{\rm kin}$
\begin{equation}
C_{\rm kin} = K_X-2G_{3 \phi} 
 + \ddot \phi (2 G_{3X} + \dot \phi ^2 G_{3XX}) + \dot \phi ^2 G_{3 \phi X} + 4H \dot \phi G_{3X} . \label{Ckin}
\end{equation}
As for $\alpha$ which describes the modification of the lensing effect, it is expressed using Eq. (\ref{Ckin})
 as~\cite{Matsumoto:2019} 
\begin{align}
\alpha = \frac{G_{4 \phi }(2G_{4 \phi}+ \dot \phi ^2 G_{3X})}
{ G_4 C_{\rm kin}+ G_{4 \phi}(-\dot \phi ^2G_{3X}+ 2G_{4 \phi})+ O(a^2H^2/k^2)} .
\label{alpha}
\end{align}
The above equation shows that $\alpha$ is non-vanishing only if $G_{4}$ is non-trivial, i.e. is $\phi$-dependent.
It also shows that $\alpha <0$ is realized only if  $G_{3X} \neq 0$ provided that we have
 $K_X >0$ and $G_4 >0$ which are satisfied for most healthy theories of gravity (see Fig.~\ref{region}). 
Thus $\alpha=0$ if $G_{4\phi} =0$, $\alpha>0$ if $G_{4\phi } \neq 0$  and $G_3 =0$,
while $\alpha$ can be positive or negative if  $G_3\neq 0$ and $G_{4\phi } \neq 0$.

In passing we note that $f(R)$ gravity theories correspond to the case \cite{de_Felice:2011,OHanlon:1972,Chiba:2003}; 
\begin{equation}
K= \frac{M_\mathrm{pl}^2}{16 \pi}(F-RF_{,R}), \quad G_3=G_5=0, 
\quad G_4 = \frac{1}{2\sqrt{8 \pi}} M_\mathrm{pl} \phi , \quad \phi 
= \frac{1}{\sqrt{8 \pi}} M_\mathrm{pl} F_{,R}\,.
 \label{frh}
\end{equation}
Because $G_3=0$, $\alpha$ is always positive in $f(R)$ gravity with $K_X>0$.
We should note that $O(a^2H^2/k^2)$ terms in Eqs.~(\ref{geff}) and (\ref{alpha}) are to be taken into account in F(R) gravity
models \cite{de_Felice:2011}. 
The case $K_X \leq 0$  with $G_3=0$ or $K_X -2G_{3\phi}\leq 0$ may be also acceptable if instabilities are absent.
Such a case is discussed in Appendix \ref{app}, but we will not consider it in the main text for simplicity.

\section{Specific examples \label{sec4}}
Let us now consider a few specific examples of Horndeski's gravity that show small but observationally 
interesting deviations from the $\Lambda$CDM model.
To compare with the $\Lambda$CDM model, we will assume 
the same amount of matter 
as that in the $\Lambda$CDM model, 
i.e. $\Omega _{m, 0} h^2$ is fixed, 
and fix the ratio $f \sigma _{8 , \Lambda CDM} / f \sigma _8=1 $ at sufficiently high redshifts, $z\geq10$. 
In what follows, to evaluate the effects of functions $K(\phi , X)$, $G_3(\phi , X)$, and $G_{4}(\phi)$,
we assume their forms as 
\begin{eqnarray}
K(\phi , X)& =& X + K_2 X^2 - \left (V_0 +V_1 \phi  +m^2\phi^2\right ), \label{ex1} \\
G_3(\phi ) &=& g\phi \,, \label{ex2} \\
G_4(\phi) &=& \frac{1}{2 \kappa ^2}\exp\left[\lambda \frac{\phi}{M_{pl}}\right]\,, \label{ex3}
\end{eqnarray}
where $K_2$, $V_0$, $V_1$, $m^2$, $g$ and $\lambda$ are constant parameters.

In this above class of models, the expressions for $\alpha$ and $G_{\rm eff}$ are
greatly simplified as
\begin{align}
\alpha &= \frac{2G_{4 \phi}^2}{G_4 C_{\rm kin} +2G_{4 \phi}^2} \, ,
\label{alphax} \\
G_{\rm eff} &= \frac{G_4 C_{kin} + 4G_{4 \phi}^2 }{16 \pi G_4(G_4 C_{\rm kin} +3G_{4 \phi}^2)} \, , 
\label{Geffx}
\end{align}
where $C_{\rm kin}$ defined in Eq. (\ref{Ckin}) is also simplified as
\begin{equation}
C_{\rm kin} = K_X - 2G_{3 \phi}\,.
\end{equation}
As $G_{4\phi}$ should be small enough to guarantee the proximity to Newton gravity
in the large scale structure formation, the above two equations imply that
$\alpha$ is small and positive if $C_{\rm kin}=O(1)$ and $>0$, and $G_{\rm eff}\approx G$.

Now we are in a position to evaluate the effect of $\alpha$ on gravitational lensing
for the same mass density distribution. Since the resulting gravitational potential is proportional
to the effective gravitational constant $G_{\rm eff}$,
the resulting gravitational lensing effect is proportional to the deflection potential given by
\begin{eqnarray}
\Psi_{\rm defl}=\frac{2+\alpha}{1+\alpha}\Psi
=\frac{2+\alpha}{1+\alpha}\frac{G_{\rm eff}}{G}\Psi_{GR}\,,
\end{eqnarray}
where $\Psi_{GR}$ is the gravitational potential of the mass distribution if gravity were GR.
Using Eqs.~(\ref{alphax}) and (\ref{Geffx}), we find
\begin{equation}
G_{\rm eff}=\frac{1}{\kappa^2 G_4}\frac{1+\alpha}{2+\alpha}G
=\frac{1+\alpha}{2+\alpha}\exp\left[-\lambda \frac{\phi}{M_{pl}}\right]G
\,,
\label{Geff}
\end{equation}
which gives
\begin{eqnarray}
\Psi_{\rm defl}=2\exp\left[-\lambda \frac{\phi}{M_{pl}}\right]\Psi_{GR}
=\exp\left[-\lambda \frac{\phi}{M_{pl}}\right]\Psi_{{\rm defl},GR}\,,
\end{eqnarray}
where $\Psi_{{\rm defl},GR}$ is the deflection potential in the case of GR.
Thus we see that the gravitational lensing effect is independent of $\alpha$ for
the same mass density distribution.
In particular, it remains essentially the same as that in GR if $\lambda\phi/M_{pl}\ll1$,
as we mentioned in section \ref{sec2}. 

\subsection{Linear potential, minimally coupled with gravity \label{4.1}}
\label{linear}
\begin{figure}
\begin{center}
\includegraphics[width=0.5\columnwidth]{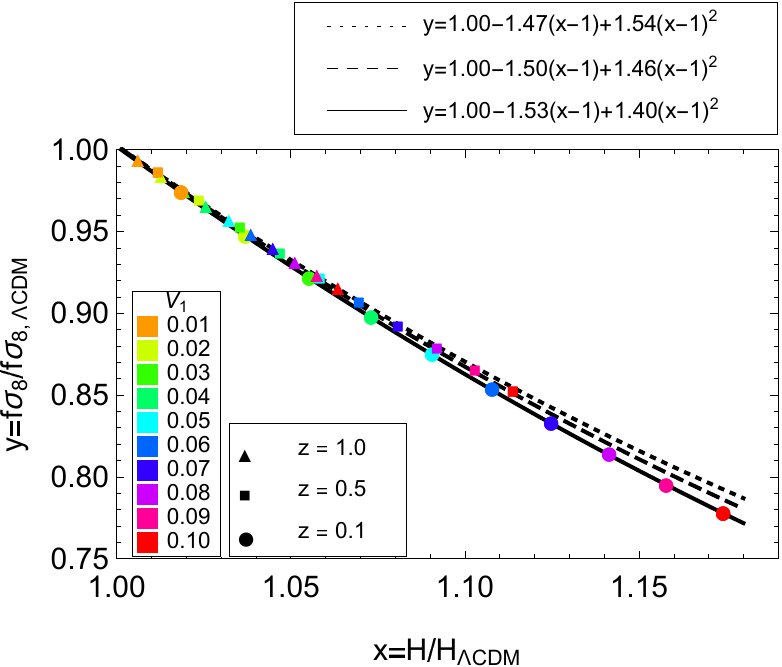}
\end{center}
\caption{$H{\rm-}f\sigma_8$ diagram in the case $K= X-(V_0 + V_1 \phi)$, $G_3=0$ and $G_{4}=const.$
	for various of $V_1$.  The initial conditions are set as
$\phi = 0.5M_{pl}$ and $\dot \phi= -2V_1/(9 H)$ at $z=10$. 
The best-fitting quadratic curves are shown for the redshifts 1.0 (dotted), 0.5 (dashed) and 0.1 (solid).}
\label{f3}
\end{figure}
First we consider the case $K=X-(V_0+V_1\phi)$ with $G_3=0$ and $G_{4}=const.$,
which represents a quintessential field \cite{Ratra:1988}. In this case, $\alpha$ always vanishes as seen from Eq.~(\ref{alpha}). 
Figure \ref{f3} shows the correlations between $f\sigma_8$ and $H$ for various values of $V_1$ at several different redshifts.
One sees that a higher Hubble rate is accompanied with a lower growth rate of the matter density perturbation,
which is similar to the case of the $w$CDM model shown in Fig.~\ref{wcdm}.
Note that $\alpha=0$ also in the $w$CDM model.
These similarities are due to the fact that both models have no non-minimal coupling with gravity.

\subsection{Flat potential, non-minimally coupled with gravity \label{4.2}}
\label{nmgrav}
\begin{figure}
\begin{minipage}[t]{0.5\columnwidth}
\begin{center}
\includegraphics[width=0.97\columnwidth]{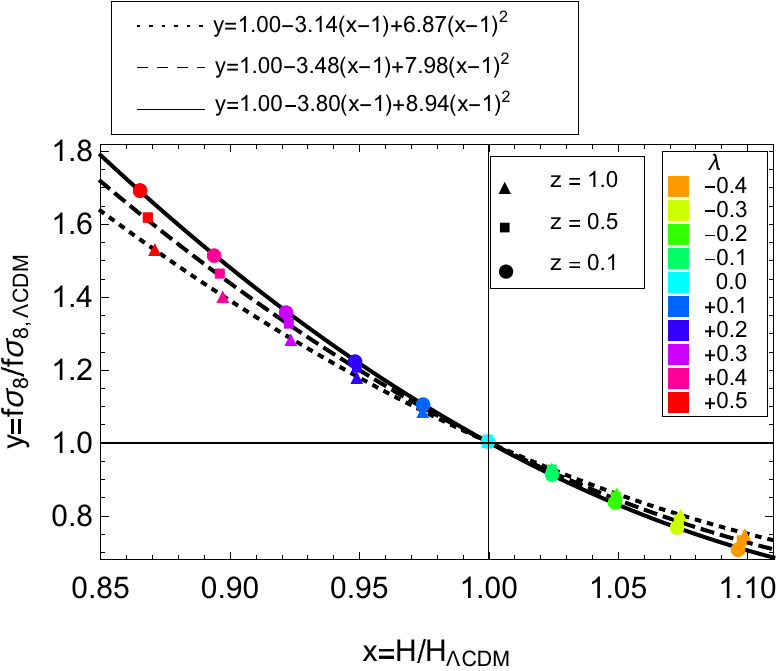}
\end{center}
\end{minipage}%
\begin{minipage}[t]{0.5\columnwidth}
\begin{center}
\includegraphics[width=0.97\columnwidth]{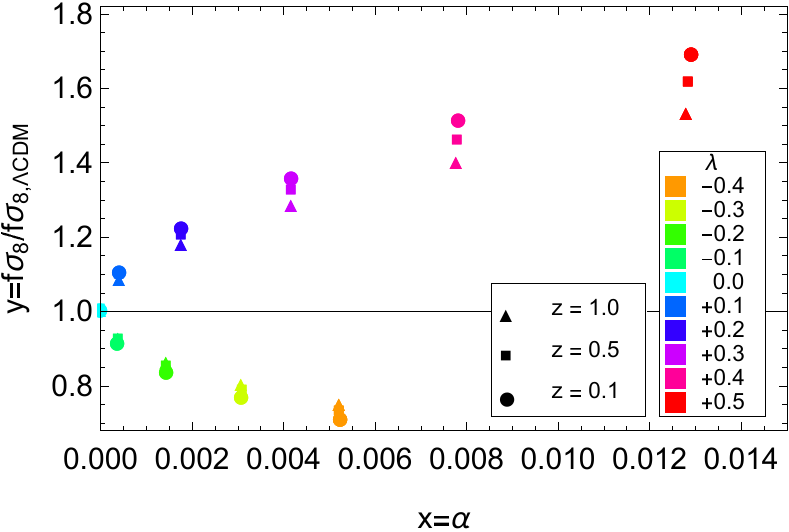}
\end{center}
\end{minipage}
\caption{
	$H-f\sigma_8$ (left) and $\alpha-f\sigma_8$ (right) diagrams
	for the case $K= X-V_0$, $G_3=0$ and $G_4=\exp[ \lambda \phi /M_{pl}]/(2 \kappa ^2)$
	for various values of $\lambda$. The initial conditions are 
$\phi= 0.5M_{pl}$ and $\dot \phi= \lambda \exp[\lambda \phi /M_{pl}]/(18 \pi H)$
at $z=10$.
In the left panel the best-fitting quadratic curves are shown for the redshifts 1.0 (dotted), 0.5 (dashed) and 0.1 (solid).
}
\label{f4}
\end{figure}

Next we consider the effect of non-minimal coupling with gravity, namely,
the case where $G_4=\exp[\lambda \phi /M_{pl}]/(2 \kappa ^2)$ with $K=X-V_0$ and $G_3=0$ (e.g., see \cite{Brans:1961,Fujii:2003}).
Figure \ref{f4} shows the correlations between $f\sigma_8$ and $H$ and between $f\sigma_8$ and $\alpha$.
The $H{\rm-}f \sigma _8$ diagram on the left is similar to that of the minimally coupled linear potential case
except for the inclinations of the fitted curves. On the other hand, the $\alpha{\rm-}f \sigma _8$ diagram 
is quite different: $\alpha$ is always non-negative as in the case of $F(R)$ gravity, see Fig.~\ref{FR}
while it vanishes in the minimally coupled linear potential case.
We also note that the ratio $f \sigma _8 /f \sigma_{8, \Lambda CDM}$ can be greater or smaller than unity depending 
on the sign of $\lambda$ in contrast to the  $F(R)$ gravity case where the ratio is always greater than unity.
The similarity and difference from the $F(R)$ gravity case can be explained by the fact that $\alpha$ does not depend 
on the sign of  $\lambda$ or $G_{4\phi}$, while $G_{\rm eff}$ may increase or decrease depending on the sign of $\lambda$.

\subsection{Quadratic potential, non-minimally coupled gravity \label{4.3}}
\label{quad}
\begin{figure}
\begin{minipage}[t]{0.5\columnwidth}
\begin{center}
\includegraphics[width=0.97\columnwidth]{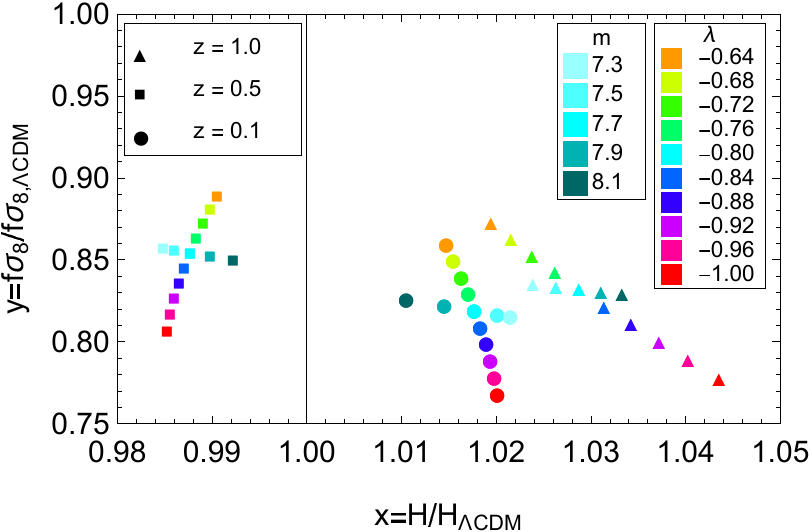}
\end{center}
\end{minipage}%
\begin{minipage}[t]{0.5\columnwidth}
\begin{center}
\includegraphics[width=0.97\columnwidth]{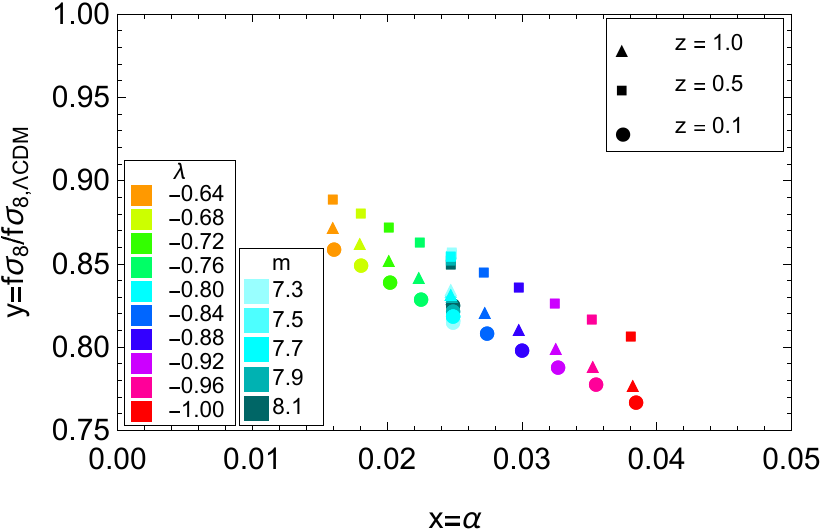}
\end{center}
\end{minipage}
\caption{Same as Fig.~\ref{f4}, but for the case $K= X-(V_0 + m^2 \phi ^2)$, $G_3=0$ and 
	$G_4=\exp[\lambda \phi /M_{pl}]/(2 \kappa ^2)$. 
	The initial conditions are 	$\phi = -0.03M_{pl}$ and $\dot \phi = 0.04H_0 M_{pl}$ at $z=10$.}
\label{f5}
\end{figure}

The third case we consider is the model which can explain the reconstructed equation-of-state parameter from recent 
observations  \cite{Matsumoto:2018}.  The Lagrangian of the model is given by 
\begin{equation}
K= X- (V_0 + m^2 \phi ^2), \quad  G_3=0,
 \quad G_4= \frac{1}{2 \kappa ^2} \mathrm{e}^{\lambda \phi / M_{pl}}. \label{vg4}
\end{equation}
This model is similar to a combination of the models considered in subsections \ref{linear} and \ref{nmgrav}.
However, it differs from such a model in that there is an oscillation of the equation of state parameter
 induced by the oscillation of the scalar field, which gives rise to an oscillatory evolution in the Hubble rate 
as can be seen in the left panel of Fig. \ref{f5}. 
We find $f \sigma _8$ in this model is always smaller than that in the $\Lambda$CDM model 
even at $z=0.5$ at which $H/H_{\Lambda CDM}<1$.
This result may be understood if we note the fact that $f(z)=d\ln D/d\ln a$, which $\sigma _8 (z)$ is proportional to,
depends non-locally on time, or it is a hysteresis effect of the oscillatory evolution in the present case.
We also note that $\alpha$ does not vary much within the range of parameters we adopted,
with the values in the range $0.02\sim0.04$. 

Actually the small positiveness of $\alpha$ is a common feature in models that satisfy $K_X -2G_{3\phi}=O(1) $
 and $G_{4\phi} \neq 0$,
provided that the background evolution does not deviate much from that in the $\Lambda$CDM model.
This may be seen by comparing Eqs.~(\ref{alphax}) and (\ref{Geffx}).
If we require $G_{4\phi}$ to be small enough to guarantee the proximity to Newton gravity,
namely, $G_{4\phi}^2/G_4\ll1$, 
these two equations implies that $\alpha$ is small and positive if $C_{kin}=O(1)$ and 
$G_{\rm eff}\approx G$.

\subsection{Non-canonical kinetic term \label{4.4}}
\label{nckin}
\begin{figure}
\begin{minipage}[t]{0.5\columnwidth}
\begin{center}
\includegraphics[width=0.97\columnwidth]{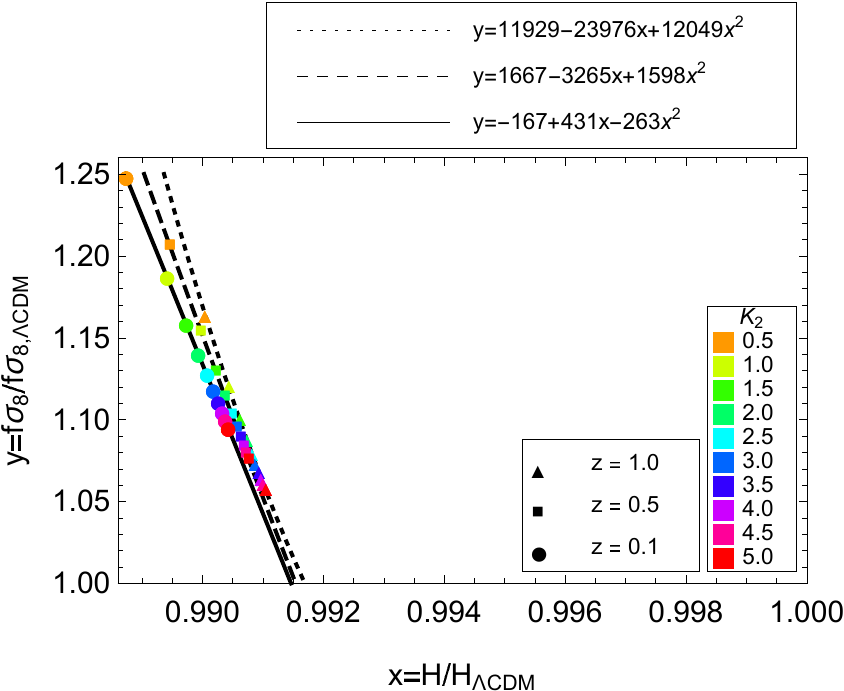}
\end{center}
\end{minipage}%
\begin{minipage}[t]{0.5\columnwidth}
\begin{center}
\includegraphics[width=0.97\columnwidth]{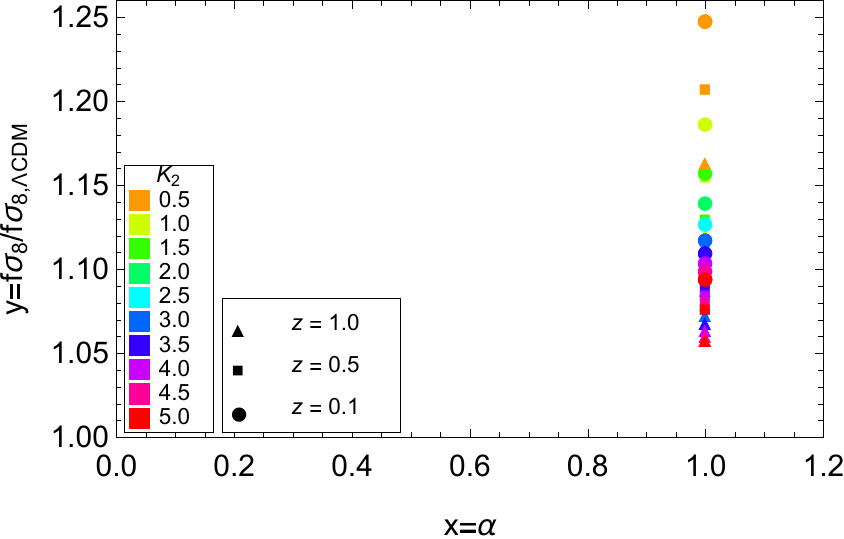}
\end{center}
\end{minipage}
\caption{Same as Fig.~\ref{f4},  but for the case $G_3 (\phi , X)=\phi /2$ and $K(\phi , X)=X+K_2 X^2 - \Lambda$
	for various values of  $K_2 /(H_0^2 M_{pl}^2)$. 
The initial conditions are	$\phi= 0.5M_{pl}$ and $\dot \phi= 0.4 H_0 M_{pl}$ at $z=10$.}
\label{f6}
\end{figure}

If we want to consider models with larger values of $\alpha$ and possibly substantial deviations 
of $G_{\rm eff}$ from Newton gravity, we need to relax the assumption $C_{kin}=K_X -2G_{3\phi}=O(1)$.
From Eq.~(\ref{alpha}), one notices that $\alpha = O(1)$ can be realized if $2G_{4 \phi}( \phi)^2\gg C_{kin}G_4$.
Since $K_X =1$ and $G_3 =0$ for a canonical scalar field, one has to resort to a non-canonical scalar field model
to achieve it. 
For definiteness, we propose the following model, 
\begin{align}
K(\phi , X)=X+K_2 X^2 - V_0\,,
\quad G_3=\frac{\phi}{2}\,,
\label{aa1}
\end{align}
which gives $C_{kin}=2K_2X$. For $K_2=O(H_0^2M_{pl}^2)$, $C_{kin}=O(\dot\phi^2/V_0)\ll 1$
for a slowly rolling scalar field. Thus this model can achieve $C_{kin}\ll1$ and hence $\alpha=O(1)$.
Here we focus on the case $K_2>0$ because the scalar field would become non-dynamical if $K_2=0$  and 
it would become a ghost if $K_2<0$.

Figure \ref{f6} shows the $H{\rm-}f\sigma_8$ and $\alpha{\rm-}f\sigma_8$ diagrams in this case
for various values of $K_2$ in units of $K_2 /(H_0^2 M_{pl}^2)$. 
As seen from the left panel, the background evolution of the Universe is 
almost same as that in the $\Lambda$CDM, with $f \sigma _8 /f \sigma _{8, \Lambda CDM}>1$.
 This behavior is similar to $F(R)$ gravity. 
The right figure shows that $\alpha =1$ is certainly realized in this case. $\alpha =1$ means that the gravitational lensing effect is enhanced by a factor $3/2$. 
We note that, for a more general form of $C_{kin}=K_X -2 G_{3\phi}$, we may achieve any value of $\alpha$ in the
range $0< \alpha <1$ while keeping the $\Lambda$CDM like background evolution intact.
We also note that $G_{\rm eff}$ would substantially deviate from the Newtonian $G$ if $G_4\approx (2\kappa^2)$
as seen from Eq.~(\ref{Geff}).

The initial value of the scalar field is chosen rather arbitrarily, except for the model discussed in Sec.~\ref{4.3}, 
but in a way that it can exhibit characteristic, qualitative features of each model. 
For example, in the case of the model considered in Sec.~\ref{4.1}, the tendency shown in Fig.~\ref{f3} remains 
the same for other choices of the initial conditions. As for Sec.~\ref{4.3}, the initial condition is chosen so that 
it reproduces the observationally constrained/indicated evolution of $w$. 
One may worry about the validity of the small scale approximation $k^2/(a^2H^2) \gg 1$ because 
some of modified gravity theories have strong dependence on the wave number $k$. 
In fact, the matter density perturbation can strongly depend on the scale in Horndeski's theory if 
there is a hierarchy between $|K(\phi X)|$, $M_{pl} H^2 |G_3(\phi ,X)|$, and $H^2 G_4$ or if there is a hierarchy between 
their derivatives. In the case of $F(R)$ gravity, 
$|K_{, \phi \phi}|\sim 1/f_{RR} \gg H^2$ should be satisfied to accord with observations and these conditions 
cause the scale dependence of the matter density perturbation. 
The reason why the scale dependence appears is that there are two limits \cite{Matsumoto:2013}: 
$k^2/(a^2H^2) \gg 1$ and $1 \gg H^2 f_{RR}$. 
The Compton wavelength is determined by balancing these two conditions, then the condition 
$k^2/(a^2H^2) \gg 1$ is superior to $1 \gg H^2 f_{RR}$ inside the Compton wavelength.  
The cases we considered in this section do not have such a hierarchy, therefore, the small scale approximation 
$k^2/(a^2H^2) \gg 1$ is always valid if we focus on a scale which is deep inside the horizon. 
\section{Conclusion \label{sec5}}
We have investigated a class of dark energy models based on Horndeski's theory of modified gravity
which exhibit small but interesting deviations from the $\Lambda$CDM model 
not only in the background evolution but also in the linear perturbation level. 
The models include the $w$CDM, $F(R)$ gravity, and four kinds of Horndeski gravity models.
To classify the properties of these models, 
we have introduced two diagrams; $H{\rm-}f \sigma _8$ and $\alpha{\rm-}f \sigma _8$ diagrams.
We have found that these two diagrams provide a useful tool to distinguish the
differences in the observational predictions of different models from each other.


\begin{figure}
\begin{center}
\includegraphics[width=0.97\columnwidth]{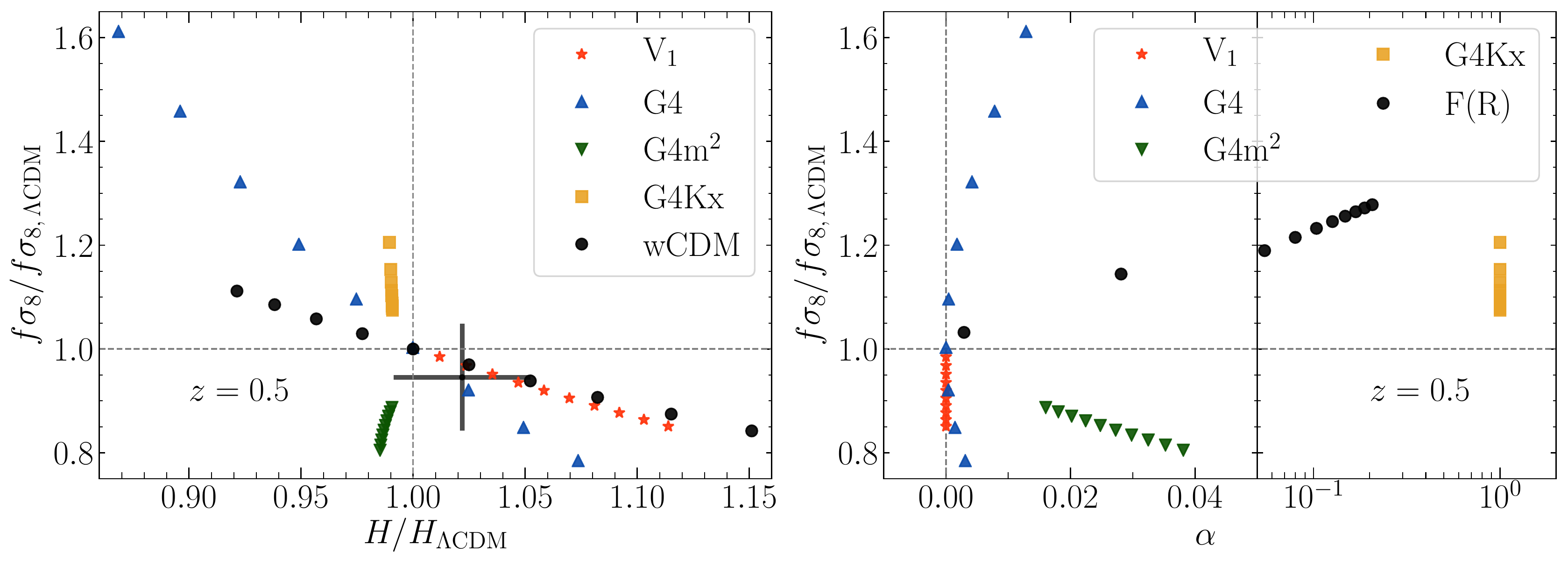}
\end{center}
\caption{Summary of the $H{\rm-}f\sigma$ (left) and $\alpha{\rm-}f\sigma_8$ (right) diagrams 
	for various dark energy models 	at redshift $z=0.5$. 
In the legend, V$_1$, G4, G4m$^2$, and G4Kx represent the cases studied in subsections \ref{linear},
\ref{nmgrav}, \ref{quad}, and \ref{nckin}, respectively. 
The $w$CDM model and $F(R)$ gravity model
are also plotted, respectively, in the $H{\rm-}f\sigma$ diagram and $\alpha{\rm-}f\sigma_8$ diagram
for comparison.  The black cross expresses on the left panel $1 \sigma$ constraint from 
the observation by Alam $et al.$ \cite{Alam:2016a}.
Note that the horizontal axis of the right panel mixes linear and logarithmic scales for clarity.}
\label{sumfig}
\end{figure}

Figure \ref{sumfig} shows the summary of our results.
 The right panel shows the $\alpha-f\sigma_8$ diagram, which exhibits
 deviations in the behavior of linear perturbations from GR, at $z=0.5$ for various models with various parameters.
The canonical scalar model with linear potential (denoted by V$_1$, discussed in section \ref{linear})  does not 
have a direct coupling to gravity, hence $\alpha =0$.
The non-minimally coupled scalar model with flat potential (denoted by G4, discussed in section \ref{nmgrav}) 
gives non-vanishing $\alpha$ but the values are too small ($\sim O(0.01)$) to be seen in the figure,
whereas the non-minimally coupled scalar model with quadratic potential (denoted by G4m$^2$, 
discussed in section \ref{quad}) .
As for the non-minimally coupled scalar model with non-canonical kinetic term 
(denoted by G4Kx, discussed in section \ref{nckin}), $\alpha =1$, that is the dynamics of the linear
perturbation is quite different from GR.

In all of the models studied in this paper, we have found $\alpha>0$. As discussed in Appendix \ref{app},
there exist theoretically acceptable models with negative $\alpha$, but they all satisfy $\alpha<-2$.
In fact, the condition $\alpha<-2$ is necessary to guarantee the positivity of $G_{\rm eff}$ as 
can be seen from Eq.~(\ref{Geff}).  Thus either $\alpha>0$ or $\alpha<-2$,
our result implies that gravity becomes effectively stronger than GR in all the models. 
In other words, if we are to measure the mass density distribution,
we would overestimate it if we use the gravitational dynamics, while we would obtain a correct estimate
if we use gravitational lensing observations. Whether this has any substantial implications to
observational data analysis is an issue to be studied.

The left panel shows the $H{\rm-}f \sigma _8$ diagram at $z=0.5$ for various models with various 
parameters. It shows the dependence of the growth rate of linear perturbations on the difference in
the background evolution of the Universe. 
We see that in all models except for the non-minimally coupled scalar model with quadratic potential 
or non-canonical kinetic term, there is a tendency that larger $H$ gives smaller $f\sigma_8$ 
in comparison with $\Lambda$CDM. In the case of the quadratic potential model (G4m$^2$),
 because of the oscillatory feature in $H$, the fact that $f\sigma_8$ are smaller for smaller $H$ 
 depends on the redshift, as well as on the choice of the model parameters. So it is difficult to discuss
  a general tendency in this model. In the case of the non-canonical kinetic term model (G4Kx),
  the dependence of $f\sigma_8$ on $H$ seems very small. This is explained by the fact that
  this model can mimic the $\Lambda$CDM model very well as long as the background evolution 
  of the Universe is concerned.

The cosmic shear power spectrum in weak lensing surveys and the redshift-space power spectrum in galaxy surveys directly constrain $\alpha$ and $f\sigma_8$, respectively. The latter observable can further probe $H$ using BAO as a standard ruler.
The blue bars denote observational $1\sigma$ error bars obtained by Alam $et~al.$ \cite{Alam:2016a}.
Since it is a $1 \sigma$ constraint, no reliable conclusion can be drawn, but it seems
that when compared to the $\Lambda$CDM model,
those models that yield slightly larger $H$ with slightly smaller $f \sigma _8$ are preferred. 
It is, however, important to note that most of the current observational analyses have been performed assuming $\Lambda$CDM and GR, a so-called consistency test. In order to test a modified gravity model, one needs to use a theoretical template of the power spectrum with the given gravity model \cite{Koyama:2009,Taruya:2014,Song:2015}. Ref. \cite{Takushima:2014} has derived an analytic formula of the matter power spectrum in real space under Horndeski's theory. Further theoretical efforts are required in order to constrain general modified gravity models on the $H-f\sigma_8-\alpha$ diagrams proposed in this paper.

 So far, observational constraints are not so severe yet to exclude any of these models.
 But eventually we will be able to exclude most of the models  as observational accuracies improve.
 The $H{\rm-}f\sigma$ and $\alpha{\rm-}f\sigma_8$ diagrams we introduced, or their variants, 
 may play an important role at such a stage.

\section*{Acknowledgments}
We would like to thank T. Namikawa for discussions in the early stage of this work.
T.~O. acknowledges support from the Ministry of Science and Technology of Taiwan under Grants 
No. MOST 106-2119-M-001-031-MY3 and the Career Development Award, Academia Sinina (AS-CDA-108-M02) 
for the period of 2019 to 2023. The work of M. S. is supported in part by JSPS KAKENHI No.~20H04727.

\appendix

\section{Horndeski's theory}
\label{Htheory}
Let us first recapitulate the action in Horndeski's theory \cite{Horndeski:1974,Deffayet:2011,Kobayashi:2011},
\begin{equation}
S_H=\sum ^5 _{i=2} \int d ^4 x \sqrt{-g} {\cal L}_i,
\end{equation}
where
\begin{align}
{\cal L}_2 &= K(\phi , X), \\
{\cal L}_3 &= -G_3(\phi , X) \Box \phi , \\
{\cal L}_4 &= G_4(\phi , X)R + G_{4 X} \left [ ( \Box \phi )^2 - (\nabla _\mu \nabla _\nu \phi)^2 \right ], \\
{\cal L}_5 &= G_5 (\phi , X)G_{\mu \nu} \nabla ^{\mu}\nabla ^{\nu} \phi - 
\frac{G_{5X}}{6} \left [ (\Box \phi)^3 - 3 (\Box \phi) (\nabla _ \mu \nabla _ \nu \phi)^2 +2 (\nabla _ \mu \nabla _ \nu \phi)^3 \right ]. 
\end{align}
Here, $K$, $G_3$, $G_4$, and $G_5$ are generic functions of 
$\phi$ and $X=- \partial _\mu \phi \partial ^\mu \phi /2$, and the subscript $X$ means a derivative with respect to $X$. 
The total action is the sum of $S_H$ and the matter action $S_\mathrm{matter}$.

The Friedman equations are given by \cite{Kobayashi:2011}
\begin{equation}
\rho _\mathrm{matter} + \sum ^5 _{i=2} {\cal E}_i=0 \label{FL1}, 
\end{equation}
where
\begin{align}
{\cal E}_2 &= 2XK_X - K, \\
{\cal E}_3 &=6X \dot \phi HG_{3X} -2X G_{3 \phi}, \\
{\cal E}_4 &= -6H^2 G_4 +24 H^2 X (G_{4X}+XG_{4XX})-12HX \dot \phi G_{4 \phi X}
-6H \dot \phi G_{4 \phi}, \\
{\cal E}_5 &= 2H^3 X \dot \phi (5 G_{5X}+ 2XG_{5XX}) -6 H^2 X (3G_{5 \phi} +2XG_{5 \phi X}), 
\end{align}
and 
\begin{equation}
p _\mathrm{matter} + \sum ^5 _{i=2} {\cal P}_i=0 \label{FL2}, 
\end{equation}
where
\begin{align}
{\cal P}_2 &= K, \\
{\cal P}_3 &= -2X \Big (  G_{3 \phi}+ \ddot \phi G_{3X} \Big), \\
{\cal P}_4 &= 2(3H^2 +2 \dot H) G_4 -4 H^2 X \bigg ( 3+ \frac{\dot X}{HX} +2 \frac{\dot H}{H^2} \bigg ) G_{4X} \nonumber \\
&-8HX \dot X G_{4XX} +2(\ddot \phi +2H \dot \phi ) G_{4 \phi }+4XG_{4 \phi \phi} + 4X(\ddot \phi -2H \dot \phi ) G_{4 \phi X}, \\
{\cal P}_5 &= -2X(2H^3 \dot \phi +2H \dot H \dot \phi +3H^2 \ddot \phi) G_{5 X}-4 H^2 X^2 \ddot \phi G_{5XX} \nonumber \\
&+4HX(\dot X -HX)G_{5 \phi X} +2 H^2 X \bigg ( 3+ 2\frac{\dot X}{HX} + 2 \frac{\dot H}{H^2} \bigg )G_{5 \phi} 
+4HX \dot \phi G_{5 \phi \phi}. 
\end{align}
Here, $H=\dot a /a$ is the Hubble rate function and the dot means derivative with respect to time and
 $\rho _\mathrm{matter}$ and $p _\mathrm{matter}$ are the matter energy density and the pressure, respectively. 
The equation of motion of the scalar field is given by varying the action with respect to $\phi (t)$: 
\begin{equation}
\frac{1}{a^3} \frac{d}{dt} (a^3 J) = P_\phi, \label{FE}
\end{equation}
where
\begin{align}
J=& \dot \phi K_X + 6HXG_{3X} -2 \dot \phi G_{3 \phi} + 6H^2 \dot \phi (G_{4X}+2XG_{4XX})-12HXG_{4 \phi X} \nonumber \\
&+ 2H^3 X(3G_{5X} + 2XG_{5XX}) -6H^2 \dot \phi (G_{5 \phi} + XG_{5 \phi X}),
\end{align}
\begin{align}
P_{\phi} =& K_\phi -2X (G_{3 \phi \phi} + \ddot \phi G_{3 \phi X}) + 6(2H^2 + \dot H) G_{4 \phi} +6H(\dot X +2HX) G_{4 \phi X} \nonumber \\
&- 6H^2 X G_{5 \phi \phi} + 2H^3 X \dot \phi G_{5 \phi X}.
\end{align}
Equations (\ref{FL1}), (\ref{FL2}), and (\ref{FE}) control the background evolution of the Universe. 
In the same manner as the quintessence model, Eqs.~(\ref{FL2}) and (\ref{FE}) are equivalent when Eq.~(\ref{FL1}) holds. 
Equations (\ref{FL1}) and (\ref{FL2}) can be rewritten in the well-known form
\begin{align}
3H^2 = \kappa ^2 (\rho _\mathrm{matter} + \rho _\phi), \\ 
-3H^2 -2 \dot H = \kappa ^2 (p_\mathrm{matter} + p_\phi),
\end{align}
where we defined $\rho _{\phi}$ and $p _\phi$ as 
\begin{equation}
\rho _{\phi} \equiv \sum ^5 _{i=2} {\cal E}_i + \frac{3H^2}{\kappa ^2}, \qquad 
p_{\phi} \equiv \sum ^5 _{i=2} {\cal P}_i - \frac{1}{\kappa ^2} (3H^2 + 2 \dot H). \label{rhop}
\end{equation}
These equations define the effective energy density and effective pressure, respectively. 
As for the matter part, we may set $p_\mathrm{matter}=0$ as the pressures of both baryons and cold dark matter are
negligible.

\section{Possibility of  $\alpha<0$\label{app}}
As mentioned in Sec.~\ref{sec4}, a negative $\alpha$ can be realized in the case 
$G_{4 \phi} \neq 0$ and $G_{3X} \neq 0$. To see this, we recapitulate the expression for $\alpha$,
\begin{eqnarray}
\alpha=\dfrac{G_{4 \phi }(2G_{4 \phi}+ \dot \phi ^2 G_{3X}) }
{ G_4 \left[K_X-2G_{3 \phi}+ \ddot \phi (2G_{3X}+ \dot \phi ^2 G_{3XX})+ \dot \phi ^2 G_{3 \phi X}\right] }\,.
\end{eqnarray}
Hence we have $\alpha<0$  if we consider a model with $G_{4\phi}(2G_{4 \phi} + \dot \phi ^2 G_{3X})<0$,
which may be realized by making $\dot \phi ^2 G_{3X}$  the same order of $G_{4 \phi}$.
A simplest model would be to set $G_3=-X/m^3$ and $K_X=1$ with $m^3\lesssim H_0^2M_{pl}$.
But we have not checked if this model could give an observationally viable model or not.

Another possibility is to consider $C_{kin}= K_X -2G_{3 \phi} \ll 1$.
In this case it seems there are many ways to realize a negative $\alpha$.
Here let us consider the case $C_{kin} =K_X -2G_{3 \phi} < 0$.
In this case, we should take care of no ghost and no gradient instability conditions. 
No ghost condition and the condition $c_s^2 \geq 0$ are expressed as \cite{Matsumoto:2018,De_Felice:2012}
\begin{eqnarray}
&&K_X+\dot \phi ^2 K_{XX}-2G_{3 \phi}- \dot \phi ^2 G_{3 \phi X}
\nonumber\\
&&\qquad +3H \dot \phi (2G_{3X}+ \dot \phi ^2 G_{3XX})
+\frac{3}{4}\frac{(2G_{4 \phi}- \dot \phi ^2 G_{3X})^2}{G_4} >0\,,
\label{s1}\\
&&G_4 C_{kin} + 4G_{4 \phi}^2 -\frac{1}{4}(2G_{4 \phi}- \dot \phi ^2 G_{3X}) ^2 \geq 0\,, 
\label{s2}
\end{eqnarray}
where $G_4>0$ is assumed to prevent the instability in the tensor perturbation. 

A simple example to realize  a negative $\alpha$ without instabilities is the case 
$G_3=0$, $K(\phi,X)=-c\,X$, $c>0$, and $G_4=\mathrm{e} ^{\lambda \phi /M_{pl}}/(2 \kappa ^2)$. 
In this case, the stability conditions (\ref{s1}) and (\ref{s2}) are expressed by a single equation; 
\begin{align}
-c\,G_4 + 3G_{4 \phi}^2 >0, 
\end{align}
which is re-expressed as
\begin{equation}
8\pi c <\frac{3}{2}\lambda ^2 \exp[\lambda \phi /M_{pl}]\,. 
\label{stab}
\end{equation}
The expressions for $\alpha$ and $G_{\rm eff}$ are given by
\begin{align}
\alpha 
&= \frac{2G_{4 \phi}^2}{-c\,G_{4 \phi}+2G_{4 \phi}^2}
= \frac{ \lambda ^2e^{\lambda \phi / M_{pl}}}{-8 \pi c+ \lambda ^2e^{ \lambda \phi / M_{pl}}}\\
G_{\rm eff} &= \frac{-c\,G_4+ 4G_{4 \phi}^2}{16 \pi G_{4}(-c\,G_4 + 3G_{4 \phi}^2)} 
=  \exp[-\lambda \phi /M_{pl}]
\frac{16 \pi c- 4\lambda ^2 {e}^{ \lambda \phi /M_{pl}}}{16 \pi c-3 \lambda ^2 {e}^{\lambda \phi /M_{pl}}} G
\end{align}
We note that this gives the same expression for $G_{\rm eff}$ as the one obtained for the models discussed in
section \ref{sec4} when expressed in terms of $\alpha$,
\begin{eqnarray}
G_{\rm eff}=\frac{2+2\alpha}{2+\alpha}\exp[-\lambda\phi/M_{pl}]G\,.
\end{eqnarray}
The stability condition (\ref{stab}) implies that $\alpha$ may take the value in the range,
$- \infty < \alpha <-2$ and $ 0 \leq \alpha < \infty$. The stability also guarantees the positivity of $G_{\rm eff}$..
A negative $\alpha$ can be realized for 
$\lambda ^2 {e} ^{\lambda \phi /M_{pl}}<8\pi c < 3 \lambda ^2 {e} ^{\lambda \phi /M_{pl}}/2$.
Both for negative and positive values of $\alpha$, the effective gravitational force
becomes twice as strong as GR in the limit of large $|\alpha|$.
\bibliographystyle{JHEP}

\providecommand{\href}[2]{#2}\begingroup\raggedright\endgroup

\end{document}